\documentclass[twocolumn,twoside,preprintnumbers,amsmath,amssymb,showkeys]
	      {revtex4}

\usepackage{epsfig}
\usepackage{graphicx}
\usepackage{fancyhdr}
\usepackage{pslatex}

\def\pip{$\pi^+$}
\def\pim{$\pi^-$}
\def\Rout{$R_{\rm out}$}
\def\Rside{$R_{\rm side}$}
\def\Rlong{$R_{\rm long}$}
\def\sg{$\sigma_{\rm G}$}

\def\pt{$p_t$}

\def\Mt{m_{\rm t}}
\def\ee{e$^+$e$^-$}
\def\runinfo{The data are preliminary and come from central Pb+Au collisions 
at 158~GeV per nucleon.}
\def\qpar{$q_{\parallel}$}
\def\qper{$q_\perp$}
\def\degrees{$^{\rm o}$}

\begin{document}
\title {Pion-pion and pion-proton correlations - new results from 
CERES\footnote{Based on the Ph.D. thesis work of Dariusz Anto\'nczyk, 
Technical University Darmstadt, 2006}}

\author{Dariusz Anto\'nczyk and Dariusz Mi\'skowiec for the CERES 
Collaboration}

\affiliation{Gesellschaft f\"ur Schwerionenforschung mbH, 
Planckstr. 1, 64289 Darmstadt, Germany}

\begin{abstract}
Results of a new two-particle correlation analysis of central 
Pb+Au collision data at 158 GeV per nucleon are presented. 
The emphasis is put on pion-proton correlations and on the 
dependence of the two-pion correlation radii on the azimuthal 
emission angle with respect to the reaction plane. 

\end{abstract}

\vskip -1.35cm

\maketitle

\thispagestyle{fancy}

\setcounter{page}{1}

\bigskip

\section{EXPERIMENT AND DATA ANALYSIS}
\label{sec:intro}

\begin{figure}[b]
\vspace{-5mm}
\hspace*{-2mm}
\rotatebox{270}{\scalebox{0.33}{\includegraphics{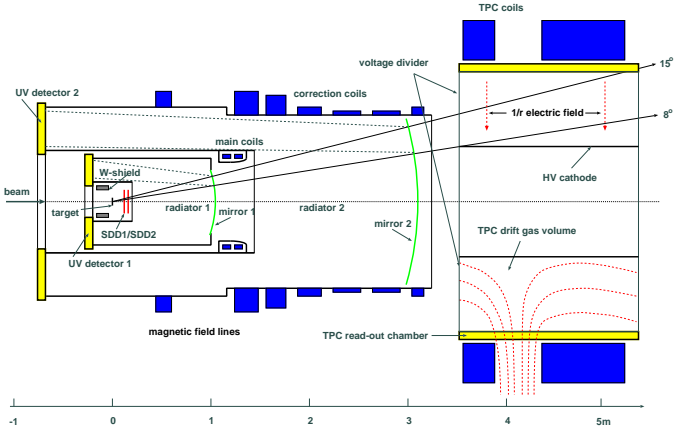}}}
\caption{
Upgraded CERES setup in 2000. 
The apparatus has a cylindrical symmetry. 
The beam enters from the left. 
The silicon detectors (SDD) give tracking and vertex reconstruction, 
and the Ring Imaging Cherenkov detectors (RICH) electron identification. 
The results presented here are mostly based on the momentum (and energy 
loss) measurements performed with the Time Projection Chamber (TPC).}
\label{fig:setup}
\end{figure}

CERES is a dilepton experiment at the CERN SPS, known for its observation 
of enhanced production of low mass \ee\ pairs in collisions between heavy 
nuclei \cite{ceres-sau-pbau}. 
The upgrade of CERES in 1997-1998 by a radial Time Projection Chamber (TPC) 
allowed to improve the momentum resolution and the particle identification 
capability while retaining the cylindrical symmetry (Fig.~\ref{fig:setup}). 
The TPC also opened the possibility of measuring hadrons. 
The upgraded experiment has been described in detail elsewhere \cite{ceres}.

\begin{figure}[b]
\rotatebox{-90}{\includegraphics[width=6.5cm]{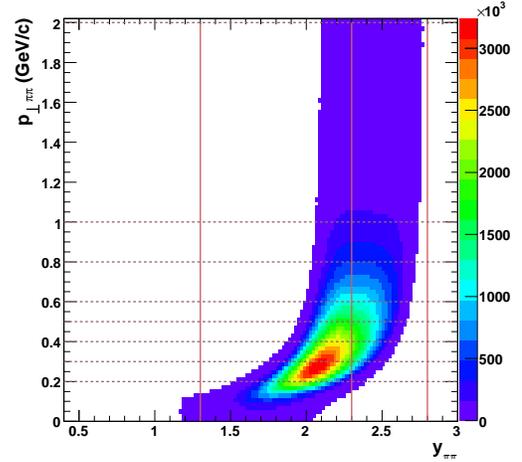}}
\rotatebox{-90}{\includegraphics[width=6.5cm]{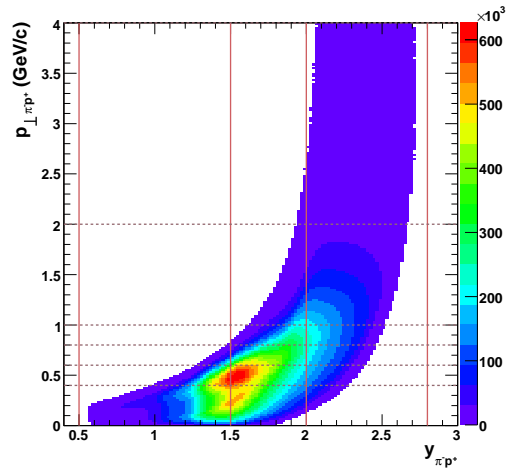}}
\caption{
Distributions of the analyzed two-pion (top) and pion-proton (bottom) pairs. 
The midrapidity is $y_B$/2=2.91.}
\label{fig:acceptance}
\end{figure}

\begin{figure}[t]
\hspace*{-5mm}
\rotatebox{-90}{\includegraphics[height=9.5cm]{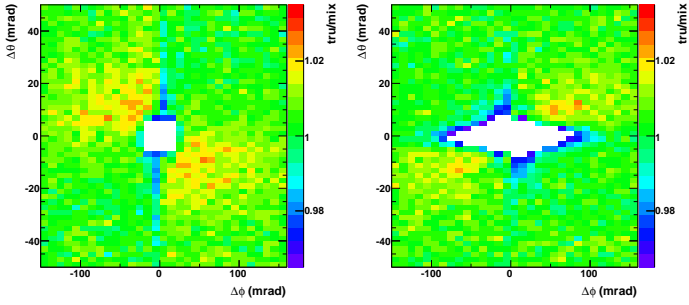}}
\caption{
Track reconstruction efficiency in the case of the magnetic field bringing the 
tracks apart from (left) or closer to (right) each other. The two topologies 
were dubbed ``sailor'' and ``cowboy'', respectively, and required different 
two-track separation cuts $\Delta \phi$: 
38-45~mr for sailor and 90-140~mr for cowboy, depending on the transverse 
momentum. The $\Theta$ separation cut was 8-9~mr.}
\label{fig:sailor-cowboy}
\end{figure}

The measurement of central Pb+Au collisions at the maximum SPS energy 
of 158 GeV per nucleon in the fall of 2000 was the first run of the fully 
upgraded CERES and at the same time the last run of this experiment. 
About 30 million Pb+Au collision events at 158 GeV per nucleon were collected, 
most of them with centrality within the top 7\% of the geometrical cross 
section \sg\ =~6.94~b. 
Small samples of the 20\% and the minimum bias collisions, as well as a short 
run at 80 AGeV, were recorded in addition. 
The first two-particle correlation analysis performed on a subset of these 
data resulted, among others, in an improved procedure to account for the 
Coulomb interaction 
\cite{heinz-pap} and a new postulate of an universal freeze-out criterion 
\cite{heinz-let}. 
The current analysis features a better momentum resolution 
\begin{equation}
\frac{\Delta p}{p} = 2\% \oplus 1\% \cdot p/{\rm (GeV\!\!/\!c)} ,
\end{equation}
a better understanding of the two-track resolution, and was performed on the 
full data set. 
The $n$(\pt,$y$) distribution of the analyzed pairs is shown in 
Fig.~\ref{fig:acceptance}.
The two-track resolution cuts applied to the true pairs and to the pairs 
from event mixing were different for the two possible track pair topologies 
(Fig.~\ref{fig:sailor-cowboy}). 
It should be noted that the required two-track cuts depended somewhat on the 
quality cuts applied to single tracks: the higher number of hits required 
for single tracks, the more pairs were lost because of the finite two-track 
resolution. 
The two-pion analysis was performed in the longitudinally co-moving frame 
(LCMS) defined by the vanishing $z$-component of the pair momentum. The 
momentum difference in this frame, {\bf q} = {\bf p}$_2$-{\bf p}$_1$, was 
decomposed into the ``out'', ``side'', and ``long'' components following the 
Bertsch-Pratt convention, with $q_{\rm out}$ pointing along the pair 
transverse momentum and $q_{\rm long}$ along the beam. The non-identical 
correlations were analyzed in the pair c.m.s., the frame in which the pair 
momentum is zero. The two components there, \qpar\ and \qper , were defined 
such that \qpar\ was along the pair transverse momentum, i.e. \qpar\ was equal 
to $q_{\rm out}$ if the latter is calculated in the pair c.m.s.

\section{Two-pion correlations}
\label{sec:two-pion}
\begin{figure}[b]
\hspace*{-3mm}\rotatebox{-90}{\includegraphics[height=9.2cm]{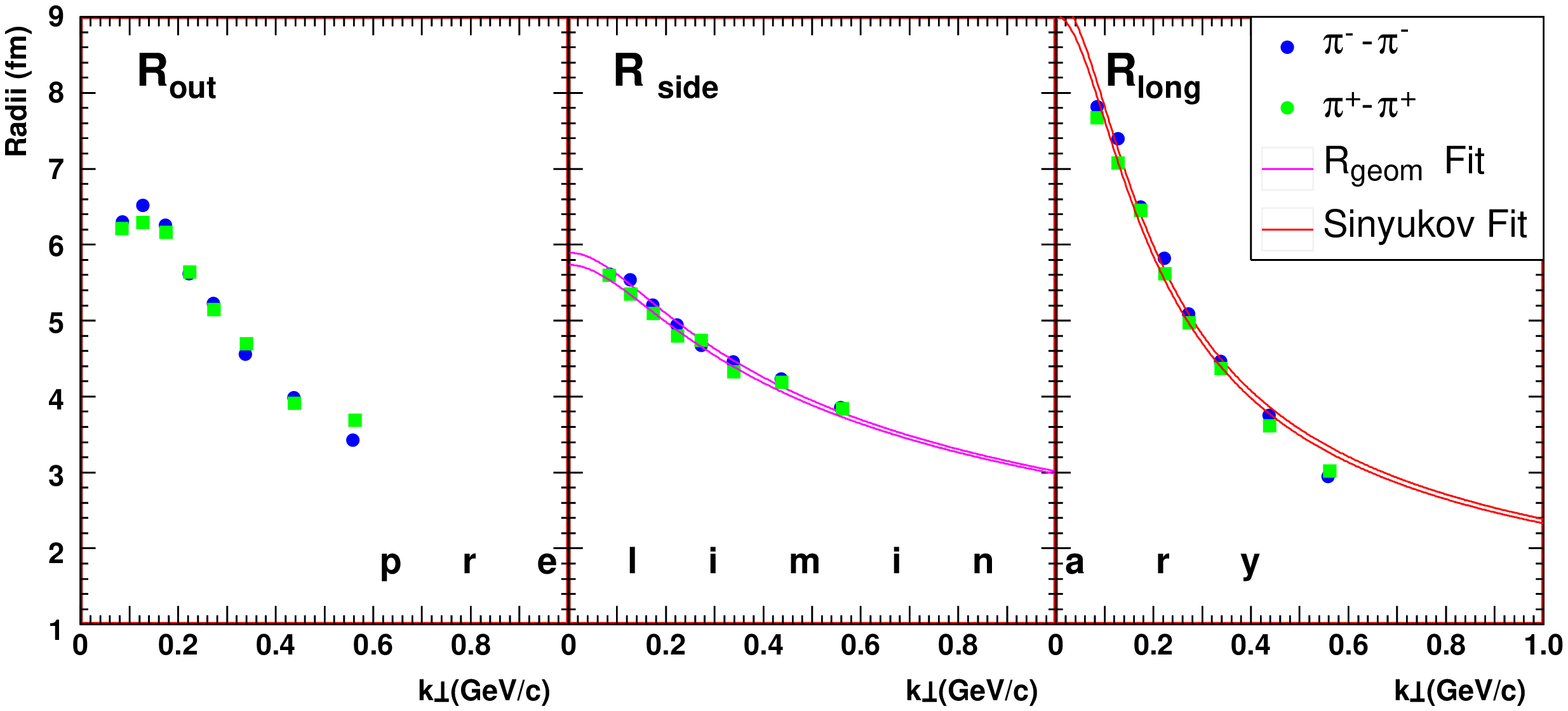}}
\hspace*{-3mm}\rotatebox{-90}{\includegraphics[height=9.2cm]{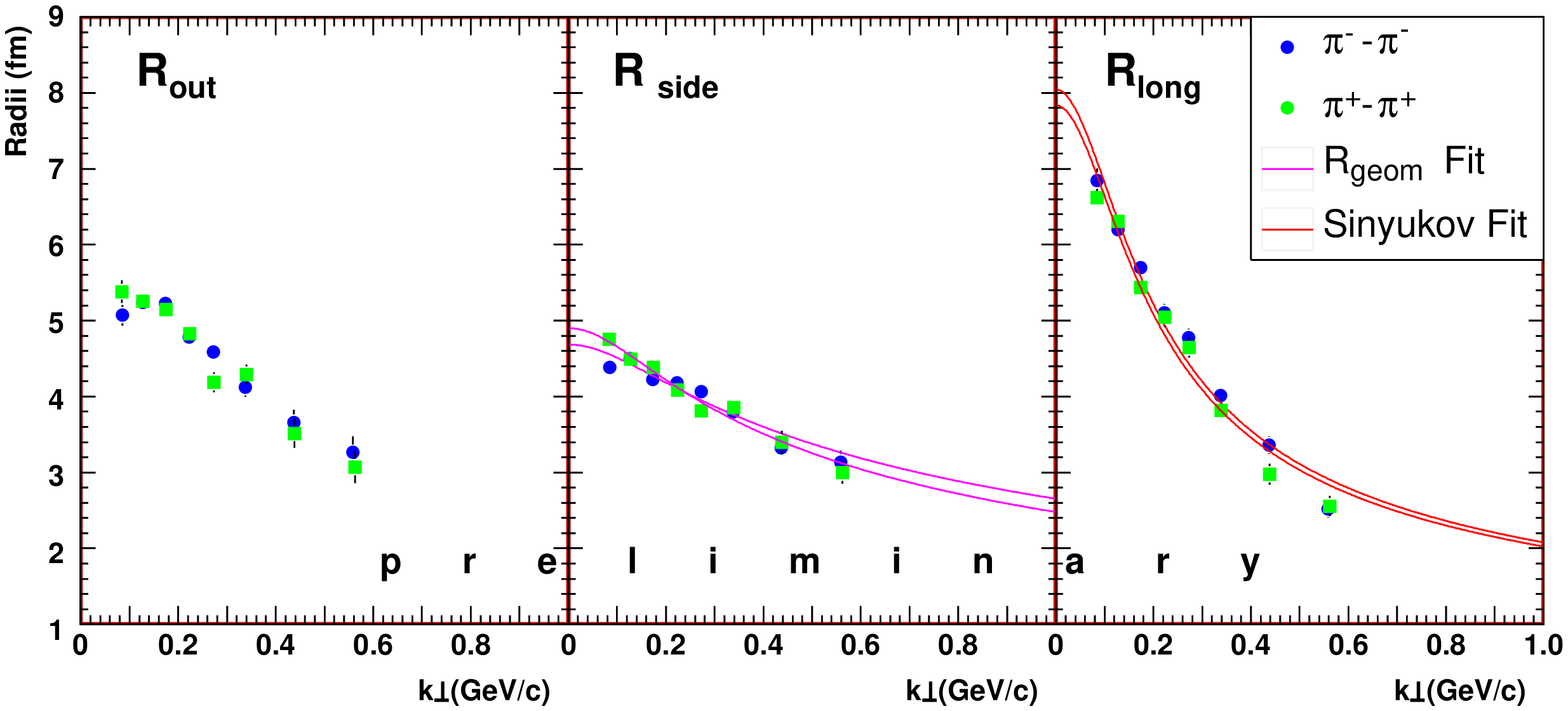}}
\caption{
Transverse momentum dependence of the source radii obtained from the 
\pim\pim\ and \pip\pip\ correlation analysis. The top and the bottom 
panels show the most and the least central bins (0-2.5\% and 15-35\% 
of \sg), respectively. The $k_\perp$ shown in the abscissa is the 
pion pair transverse momentum divided by two. \runinfo}
\label{fig:rpt}
\end{figure}

\begin{figure*}[t]
\hspace*{-5mm}
\rotatebox{-90}{\includegraphics[height=18cm]{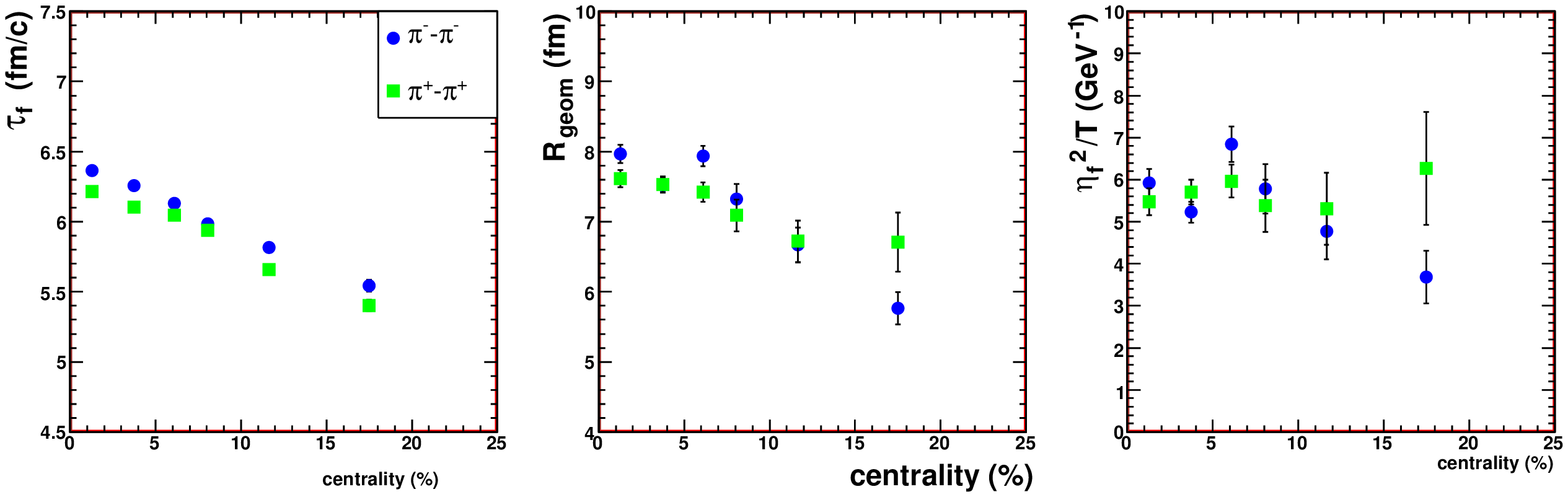}}
\caption{
Centrality dependence of the pion source parameters obtained from fits 
(\ref{eq:rsidept}) and (\ref{eq:rlongpt}). \runinfo\ }
\label{fig:hbtcent}
\end{figure*}

The \pim\pim\ and \pip\pip\ correlation functions, defined as the 
three-dimensional distributions of pion pairs from the same event 
$n(${\bf q}$)$, normalized to the analogous distributions of pairs 
constructed from different events (event mixing), were fitted by 
\begin{equation}
C_2\left({\bf q}\right)= N \cdot\left\{\left(1-\lambda\right) + 
\lambda \cdot F_{c}\left(q_{inv}\right)\left[1+
\exp\left(-\displaystyle\sum_{i,j = 1}^{3} R^{2}_{ij} 
q_{i}q_{j}\right)\right]\right\}\,\,.
\label{eq:fitfunction}
\end{equation}
The normalization factor $N$ is needed because the number of pairs from 
event mixing is arbitrary. 
The correlation strength $\lambda<$1 reflects the tails of the source 
distribution caused by the pions from long-lived resonances, the finite 
${\bf q}$-resolution, 
and the contamination of the pion sample by other particle species. 
The  $R^{2}_{ij}$ fit parameters, with the indices $i$,$j$ being 
\{out, side, long\}, are related to the size of the source emitting pions 
of given momentum \cite{mike-overview} and will be called here HBT radii. 
The $F_{c}\left(q_{inv}\right)$ factor, $q_{inv}=\sqrt{-(p_2^\mu-p_1^\mu)^2}$, 
accounts for the mutual Coulomb interaction between the pions
and was calculated by averaging the nonrelativistic Coulomb wave function 
squared over a realistic source size. 
The Coulomb factor was attenuated by $\lambda$ similarly as the rest of the 
correlation function peak; the importance of this approach was demonstrated 
in \cite{heinz-pap}. 
The fits were performed by the minimum negative loglikelihood method with 
the Poissonian number of true pairs and were done separately for each pair 
(\pt,y) bin. 
The HBT radii obtained from the fit were corrected for the finite 
momentum resolution.
The correction was determined by Monte Carlo and was rather insignificant 
for \Rside\ and \Rlong ; for \Rout\ it gets as large as $\approx$ 20\% for 
the highest bin of the pair \pt. 
The obtained HBT radii show a strong \pt\ dependence (Fig.~\ref{fig:rpt}). 
The \Rside\ and \Rlong\ radii were fitted with \cite{rside,rlong}
\begin{eqnarray}
  R_{side}(p_t) &=& \sqrt{\frac{R_G^{2}}{1+\Mt \, \, 
      \eta_{f}^{2}/T}} \label{eq:rsidept}\\
  R_{long}(p_t) &=& \tau_{f}\sqrt{\frac{T}{\Mt} 
    \frac{K_{2}\left(\Mt/T\right)} {K_{1}\left(\Mt/T\right)}}
  \label{eq:rlongpt}
\end{eqnarray} with the freeze-out temperature $T$ fixed to be 120~MeV. 
The results of the fit are shown in Fig.~\ref{fig:hbtcent}. 
The obtained expansion time $\tau$ of 5.5-6.5~fm, geometrical source size 
$R_G$ of 6-8~fm, and transverse expansion rapidity $\eta_f$ of 0.7-0.8, in 
reasonable agreement with the results of the previous analysis of a subset 
of data \cite{heinz-pap}, indicate a long-living, 
longitudinally and transversally expanding pion source. 

\section{Azimuthal angle dependent HBT}
\label{sec:reaction-plane}
\begin{figure}[b]
\hspace*{-5mm}\rotatebox{-90}{\includegraphics[width=8cm]{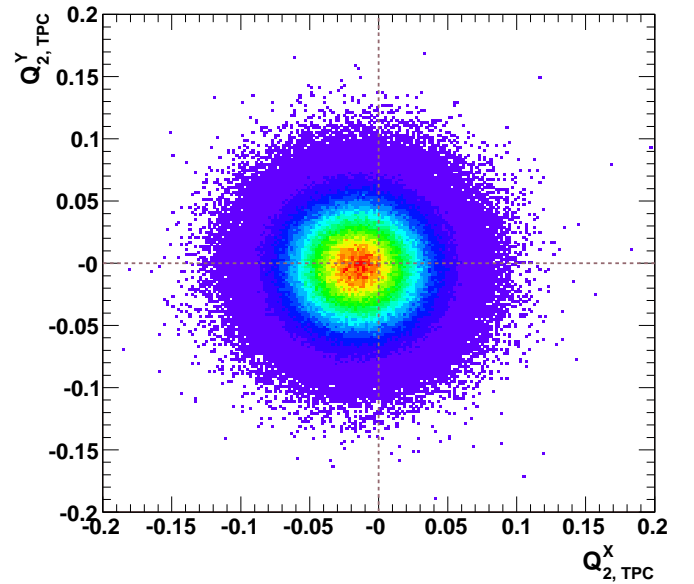}}
\caption{
The raw distribution of the $(Q^X_2,Q^Y_2)$ vector used to determine 
the azimuthal orientation of the reaction plane. 
The distribution was recentered and made round run-by-run in order 
to make the resulting distribution of the reaction plane angle uniform.}
\label{fig:qdist}
\end{figure}

A fireball created in a collision with a finite impact parameter is elongated 
in the direction perpendicular to the reaction plane. In the course of 
expansion, with the pressure gradient larger in-plane than out-of-plane, 
the initial asymmetry should get reduced or even reversed. A dependence of 
the two-pion correlations on the pair emission angle with respect to the 
reaction plane would be a signature of the source eccentricity at the 
decoupling time. 
The azimuthal angle of the reaction plane was estimated by the preferred 
direction of the particle emission aka elliptic flow. The particles were 
weighted with their transverse momentum:
\begin{eqnarray}
Q^X_2 &=& \sum_i p_t \, \cos(2\phi_i)\\
Q^Y_2 &=& \sum_i p_t \, \sin(2\phi_i) \, .
\end{eqnarray}
The raw distribution $n(Q^X_2,Q^Y_2)$ is shown in Fig.~\ref{fig:qdist}. 
The reaction plane angle was calculated (modulo $\pi$) from the calibrated 
$Q_2$ components via 
\begin{equation}
\Phi_{\rm RP} = \frac{1}{2} \arctan \left[\frac{Q^Y_2}{Q^X_2}\right] .
\end{equation}
The resolution of the so determined reaction plane angle, estimated via the 
subevevent method, was 31-38\degrees. 

\begin{figure}[h]
\hspace*{-5mm}\includegraphics[width=9cm]{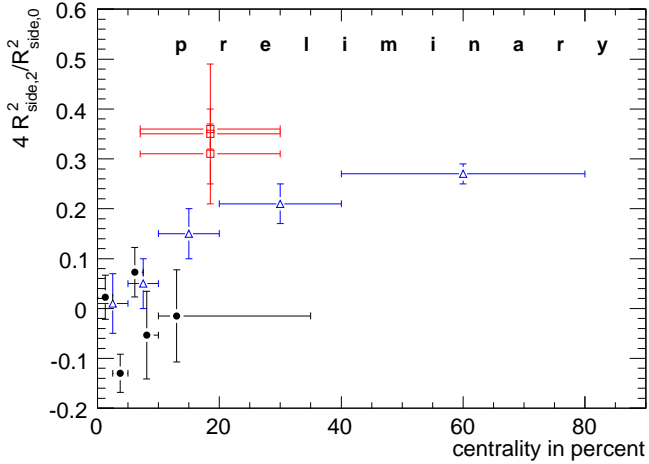}
\caption{
Azimuthal source eccentricity, represented by the normalized second Fourier 
component of \Rside$(\Phi^*)$, vs. centrality. 
The CERES result (black full dots) is much closer to zero than the 
analogous measurement at the AGS (open red squares) and RHIC (open blue 
triangles). }
\label{fig:r2}
\end{figure}

With the event plane known (within the resolution) event-by-event the 
pion pairs can be sorted into 8 bins covering $(-\pi/2,\pi/2)$ according to 
their azimuthal angle with respect to the reaction plane 
$\Phi^*=\Phi_{\rm pair}-\Psi_{\rm RP}$. 
During event mixing it was required that the two events had a similar 
reaction plane angle $\Psi_{\rm RP}$. 
The eight correlation functions were fitted as described in Section 
\ref{sec:two-pion}, and the resulting \Rout, \Rside, \Rlong, and the 
cross-terms are plotted versus $\Phi^*$. 
The squared source radii were then fitted with  
\begin{equation}
R_i^2 = R_{i,0}^2+2 R_{i,2}^2 \cos(2 \Phi^*) ,
\label{rfourier}
\end{equation}
with $i$ denoting \{out,side,long\}. While the $R_{i,0}$'s obtained coincide 
with the results of the standard HBT analysis presented in Section 
\ref{sec:two-pion} 
the second Fourier components $R_{i,2}$'s represent the eccentricity of the 
observed pion source. 
The normalized second Fourier component of \Rside\ is shown in 
Fig.~\ref{fig:r2}. 
As far as the limited centrality range allows to judge the measured \Rside\  
anisotropy is consistent with zero, in contrast to the out-of-plane 
elongated pion source observed both at the AGS \cite{ags} and at RHIC 
\cite{rhic}. 
The source eccentricity, thus, seems to be joining the exclusive club of 
heavy ion observables which behave non-monotonically with the collision 
energy, the other members being the flow and the strangeness-to-entropy ratio.

\section{Pion-proton correlations}
\label{sec:pion-proton}
\begin{figure}[t]
\hspace*{-2mm}\includegraphics[width=9cm]{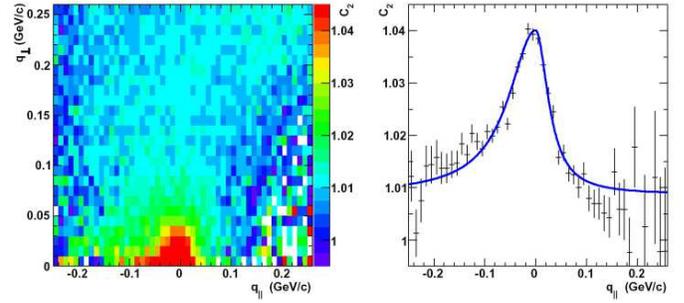}
\caption{
Two-dimensional $\pi^-$p correlation function  $C$(\qpar,\qper) (left) 
and its slice $C$(\qpar) for \qper$<$50~MeV/c (right). 
The peak asymmetry in the right hand plot indicates that the proton 
source is located at a larger radius than the pion source, or that 
protons are emitted earlier than pions. 
The asymmetry can be parametrized using Eq. (\ref{eq:AsymmetryFit}). }
\label{fig:nonident}
\end{figure}

For the top central 7\% of the geometrical cross section the high 
event statistics allows to perform the proton-pion correlation analysis. 
The shape of non-identical particle correlation functions $C({\bf q})$ 
reflects the shape of the relative source distributions $S(r_2^\mu-r_1^\mu)$. 
Particularly, a difference between the average freeze-out position or time 
of two particle species reveals itself as an asymmetry of the 
correlation function at small $q$ \cite{lednicky}. 
A two-dimensional $\pi^-$p correlation function $C$(\qpar,\qper)
and its slice $C$(\qpar) for \qper$<$50~MeV/c are shown in 
Fig.~\ref{fig:nonident}. 
The peak at low $q$ comes from the attractive Coulomb interaction. 
The peak asymmetry 
indicates that the proton source is located at a larger radius 
than the pion source, or that protons are emitted earlier than 
pions. 
The asymmetry can be conveniently parametrized by fitting a 
Lorentz curve, which happens to match the shape, 
separately to the left and to the right half of the peak, 
and taking the ratio of the two widths: 
\begin{equation}
C_{2}(q_{\parallel}) = \left\{ 
\begin{array}{ll}
N \cdot \left(1+\frac{a}{\left(q_{\parallel}/\sigma_{-}\right)^{2} + 1}
\right)\,, & q_{\parallel} < 0 \\
N \cdot \left(1+\frac{a}{\left(q_{\parallel}/\sigma_{+}\right)^{2} + 1}
\right)\,, & q_{\parallel} > 0
\end{array} \right.
\label{eq:AsymmetryFit}
\end{equation}
where $N$ is a normalization factor and $a$ is the peak amplitude. 
The asymmetry is then defined as the ratio between the two
widths $\mathcal{A}=\sigma_{-} / \sigma_{+}$.

\begin{figure}[b]
\rotatebox{-90}{\includegraphics[width=6cm]{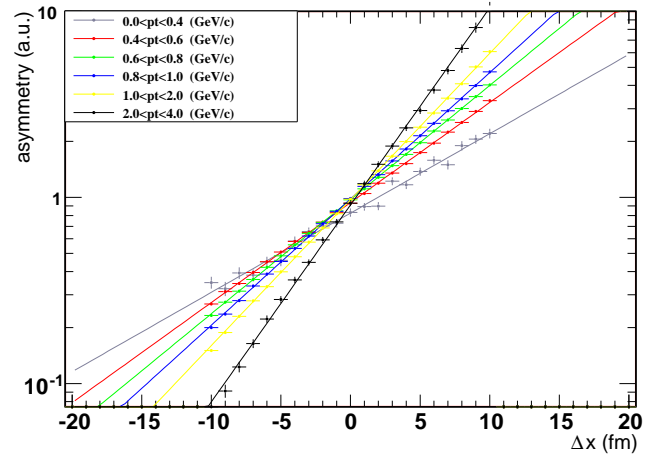}}
\caption{
Relation between the pion-proton correlation peak asymmetry 
and the displacement between the source of protons and pions, 
obtained from Monte Carlo. This relation was used to translate the 
measured asymmetries to displacements. }
\label{fig:asymmetrycalibration}
\end{figure}

\begin{figure*}[t]
\hspace*{-5mm}\rotatebox{-90}{\includegraphics[width=17cm,clip=]{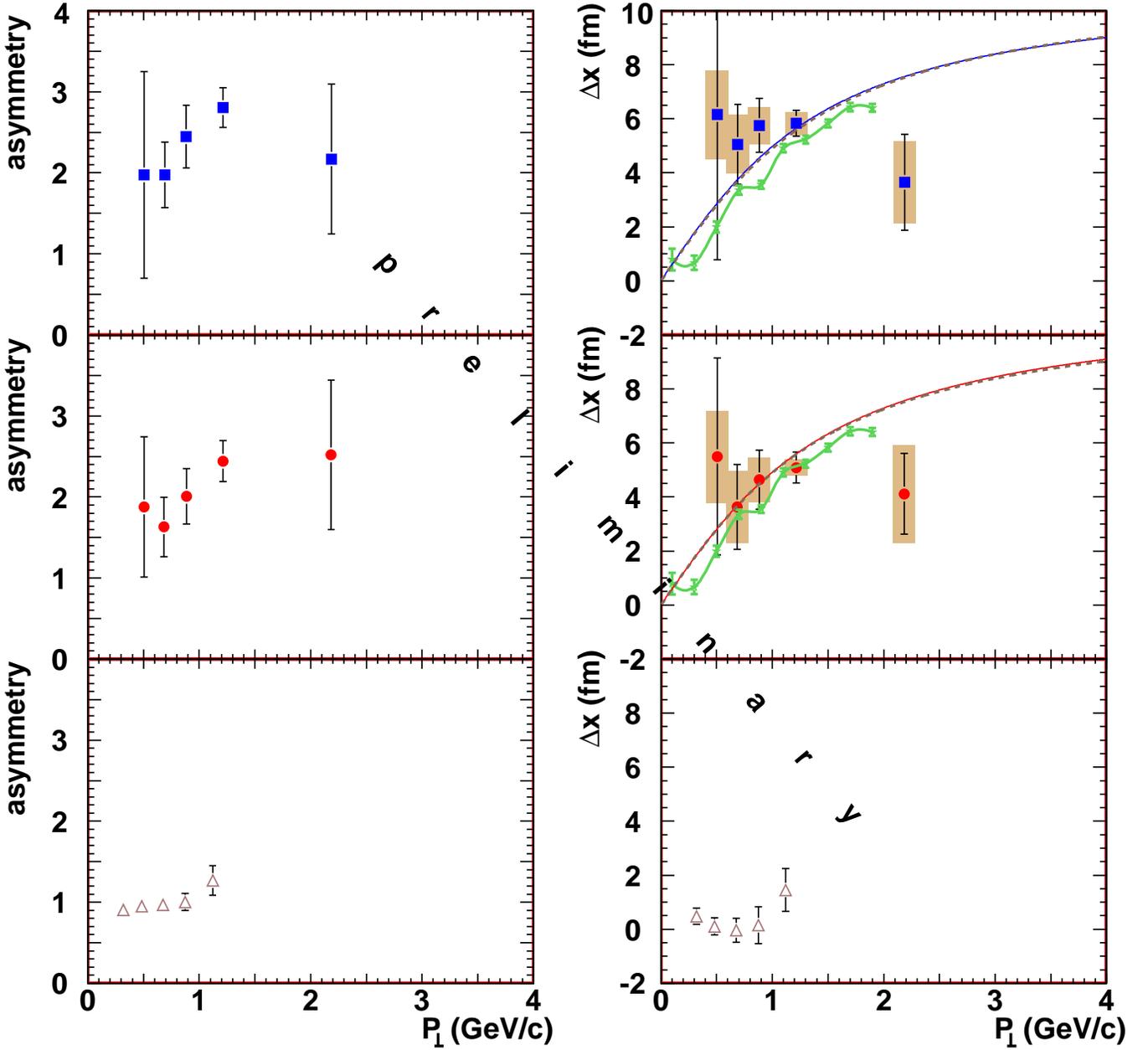}}
\caption{
Left: \pip -proton (top) and \pim -proton (center) correlation peak asymmetry, 
defined as the ratio between the two half-widths in Eq. 
(\ref{eq:AsymmetryFit}), vs. the pair transverse momentum. 
The asymmetry is larger than one which indicates that protons 
on average freeze-out at a larger radius (or earlier time) than 
pions. 
No asymmetry is visible in the \pip\-\pim\ correlations (bottom panel). 
For symmetry reasons all peak asymmetries must vanish at small \pt 's. 
Right: the source displacement deduced from the asymmetry.
The green lines represents the displacements extracted from UrQMD 
for \pt's between 0 and 2 GeV/c.  
The long lines show the fit described in text. \runinfo}
\label{fig:asymmetryresults}
\end{figure*}

The asymmetry was translated to a spatial displacement between 
the proton and pion sources using a Monte Carlo pair generator 
with realistic source sizes and the Coulomb wave function squared 
as a weighting factor for each pair. 
The relation between the two quantities is shown in 
Fig.~\ref{fig:asymmetrycalibration}. 
The resulting displacement $\Delta x$ as a function of the transverse pair 
momentum is shown in the right panel of Fig.~\ref{fig:asymmetryresults}. 
In the left panel the raw asymmetry $\mathcal{A}$ is shown. 
The asymmetry and the displacement derived from it vanish in the limit 
of small pair \pt\ as is expected for symmetry reasons. 
The displacement is rather similar to the spatial displacement seen in 
UrQMD v.1.3 \cite{urqmd} (green line). The red and blue lines represent a 
simultaneous fit to \Rside(\pt) and $\Delta x$(\pt) using, respectively, 
Eq.(\ref{eq:rsidept}) and the formula derived in \cite{deltax}: 
\begin{equation}
\langle \Delta x \rangle = \frac{R_G \, \beta_\perp \, 
  \beta_0}{\beta_0^2+\frac{T}{\Mt}}
\end{equation}
with the pair transverse mass 
\begin{equation}
    \Mt = \sqrt{ \sqrt{m_{\pi}^{2} + \frac{m_{\pi}}{m_{p}+ m_{\pi}} \cdot 
	\left(\frac{P_{\perp}}{2}\right)^{2}}
    \cdot \sqrt{m_{p}^{2} + \frac{m_{p}}{m_{\pi}+m_{p}} \cdot 
      \left(\frac{P_{\perp}}{2}\right)^{2}}}\,,
\end{equation}
and the pair transverse velocity 
\begin{figure*}[t]
\hspace*{-8mm}\includegraphics[width=18cm]{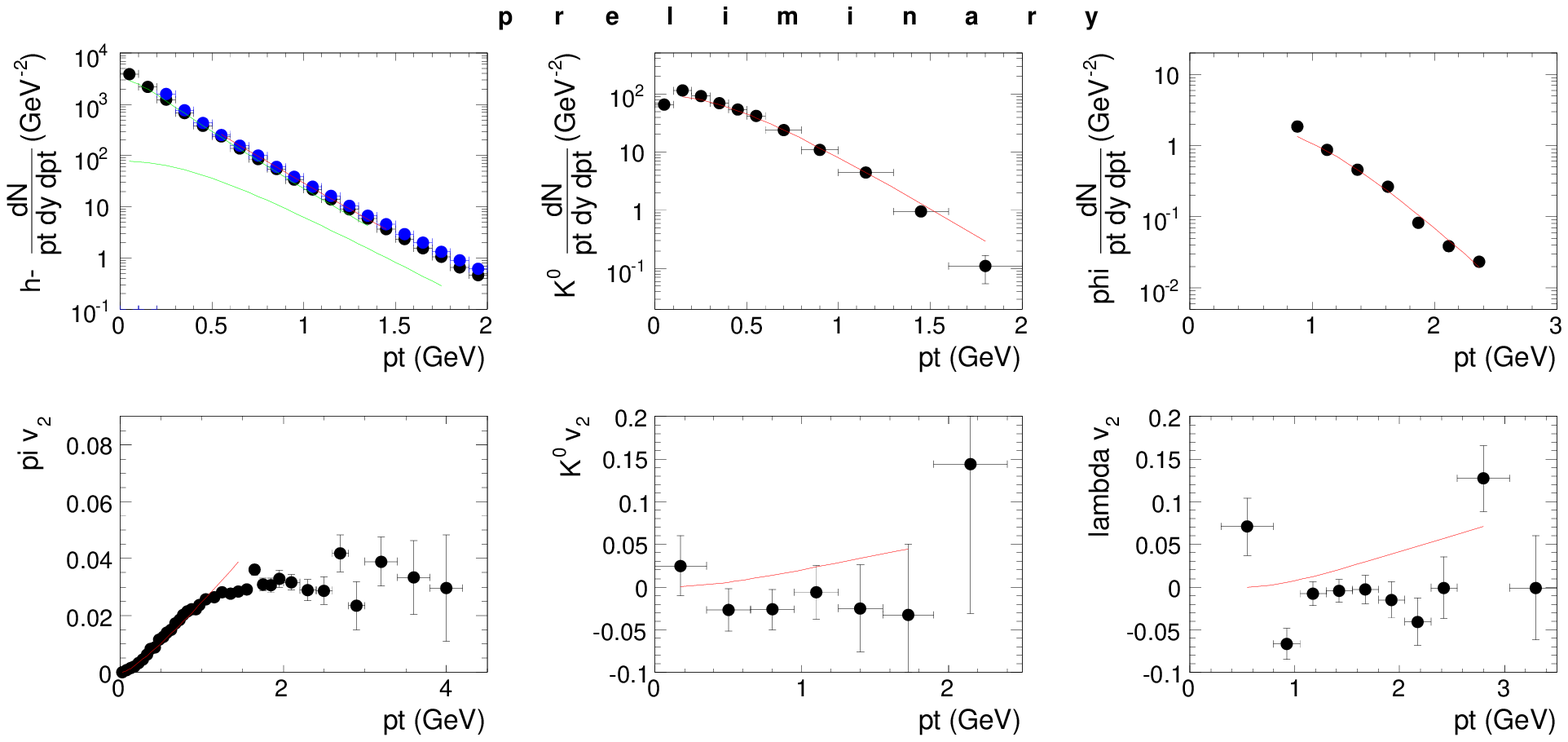}
\caption{
Transverse momentum spectra of negatively charged particles, 
K$^0$, and $\phi$-meson (top) and elliptic flow coefficient 
for pions, K$^0$, and $\Lambda$ (bottom), compared to the 
blast-wave model (red line). 
The two green lines in the top left plot represent negative pions and kaons, 
the sum of the two being the red line to be compared to the data. 
\runinfo
The black and blue points represent the negative and 
positive pions, respectively. 
}
\label{fig:blast0}
\end{figure*}
\begin{equation}
\beta_\perp = \frac{1}{\sqrt{1+\left(\frac{m_\pi+m_p}{P_\perp}\right)^2}} \, .
\end{equation}
The relation between the rapidity $\eta_f$ and the velocity $\beta_0$ of 
transverse flow, which occur in the \Rside\ and $\Delta x$ fit formulas, 
respectively, is
\begin{equation}
\eta_f = \frac{1}{2} \ln\frac{1+\beta_0}{1-\beta_0} .
\end{equation}
The fit yields a $\beta_0$ of 0.65-0.70 and a $R_G$ of about 7.5~fm. 
The fit is dominated by the \Rside\ data with their small error bars but 
it reproduces the $\Delta x$ values quite well. This indicates that the 
finite displacement between the sources of pions and protons has a similar 
origin 
as the \pt\ dependence of \Rside, namely the transverse flow. 
Other, possibly more interesting, contributions to it cannot be addressed 
with the present statistics of the data. 

\section{Blast-wave parametrization}
\begin{figure*}[t]
\hspace*{-5mm}\includegraphics[width=18cm]{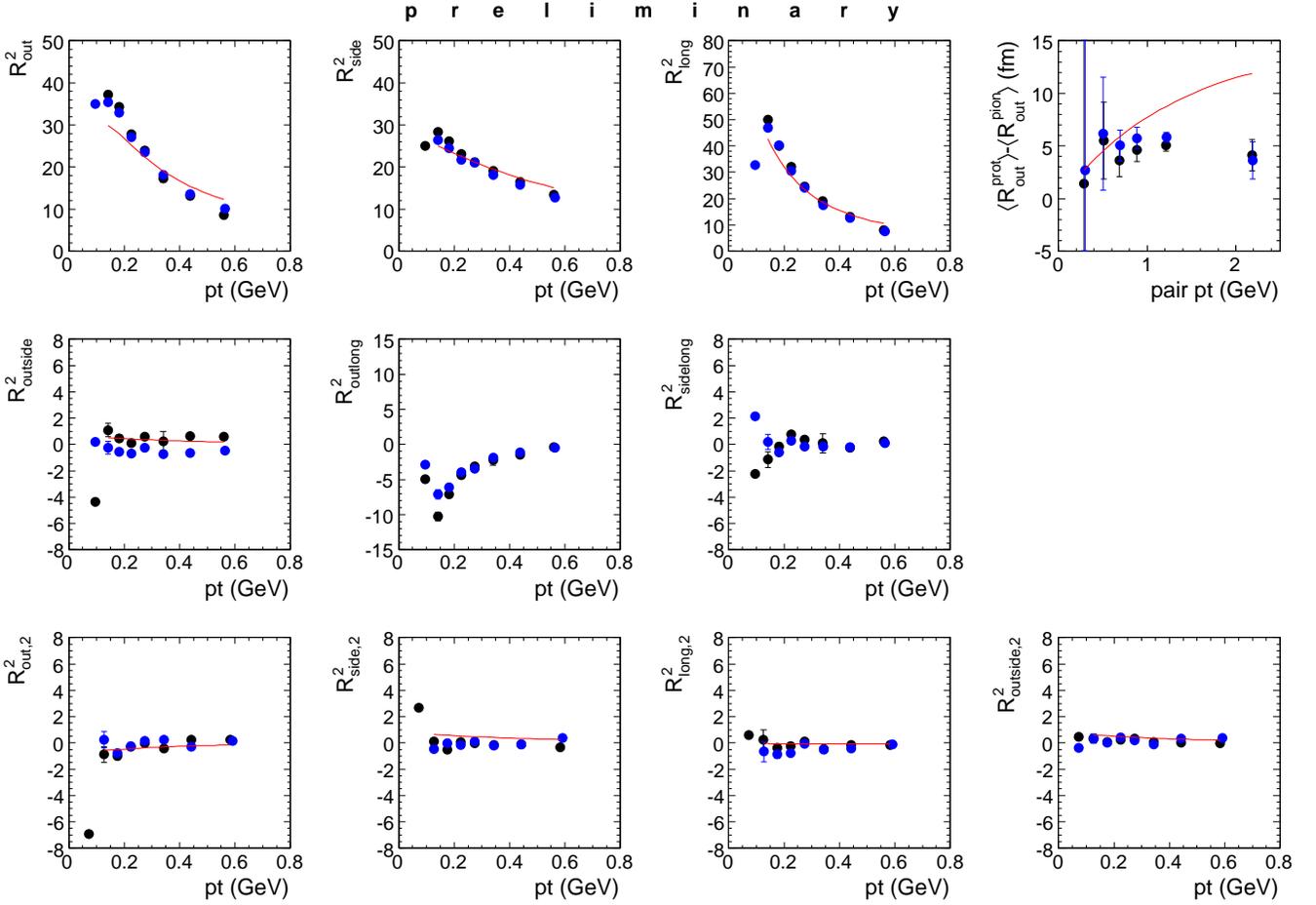}
\caption{
CERES HBT radii squared and the pion-proton displacement (top), HBT 
cross-terms (middle), and the second Fourier components of the HBT radii 
vs. emission angle with respect to the reaction plane (bottom), compared 
to the blast-wave model calculations (red line). 
\runinfo
The black and blue points represent the negative and positive pions, 
respectively. 
}
\label{fig:blast1}
\end{figure*}
While it is notoriously difficult to describe many observables 
of heavy ion collisions within the same theoretical approach, 
in recent years 
simple hydrodynamics-inpired parametrizations turn out to 
be quite successful in this aspect. 
The blast-wave model of \cite{mike-blast}, with its 8 parameters 
adjusted accordingly, nicely reproduces the transverse momentum 
spectra, the elliptic flow, and the two-particle correlations 
of CERES, including the emission angle dependence (or, rather, 
the lack of it). 
The transverse spectra and the elliptic flow coefficients from the 
top 7\% of \sg\ for several hadron species are shown along with the 
blast-wave model lines in Fig.~\ref{fig:blast0}. 
In Fig.~\ref{fig:blast1} the two-particle correlation results are 
compared to the calculations performed with the blast-wave model with the 
same parameter values. The agreement is rather good. 
Two of the eight parameters of the model were fixed: the freeze-out 
temperature $T=100$~MeV and the sharpness of the emission volume 
$a=0.01$ (sharp). The other six parameters, adjusted to the data by 
the simplex method, include the transverse flow rapidity 
$\rho$ of 0.88 and 0.85 in-plane and out-of-plane, respectively; 
the source radius $(R_x,R_y)$ of (11.26, 11.42)~fm; the emission time (i.e. 
the duration of the expansion) $\tau$ of 7.4~fm, and the duration 
of emission $\Delta \tau = 1.55$~fm/c. 
The nearly-circular source $R_x\approx R_y$ was enforced by the 
smallness of the second Fourier component of the HBT radii (bottom 
part of Fig.~\ref{fig:blast1}). 

\section{Summary}
In summary, the new high-statistics CERES hadron data allow for state 
of the art analysis of two-particle correlation data. 
The pion source anisotropy, accessible via the dependence of the 
two-pion HBT radii on the emission angle with respect to the reaction 
plane, is unexpectedly small compared to the analogous results obtained 
at lower and higher energies. 
The pion-proton correlations indicate that the proton source is located 
at a larger radius than the pion source, or that protons are emitted 
earlier than pions. 
The amount of the displacement, and its functional dependence on the 
transverse momentum of the two involved particles, corroborate the  
transverse expansion picture deduced from the behavior of the \Rside\ 
HBT radius. 
Finally, the blast-wave parametrization is able to describe simultaneously 
the transverse spectra and the elliptic flow coefficients of several 
hadron species, and the two-particle correlations. 

This work was supported by the German BMBF, the U.S. DoE, 
the Israeli Science Foundation, and the MINERVA foundation. 
The data analysis was performed by D.~Anto\'nczyk as a part of his 
Ph.D. thesis work. 
The author (DM) thanks the workshop organizers for the invitation 
and for the good and inspiring atmosphere of the meeting.


\end{document}